\documentclass[
 reprint,
 superscriptaddress,
 amsmath,amssymb,
 aps,
 prmaterials,
 nofootinbib
]{revtex4-2}

\usepackage{graphicx}
\usepackage{dcolumn}
\usepackage{bm}
\usepackage{mhchem}

\begin{document}

\title{Robust and efficient calculation of activation energy by automated path search and density functional theory}

\author{Koki Ueno}
\affiliation{Technology Innovation Division, Panasonic Corporation, Osaka 571-8508, Japan}
\author{Kazuhide Ichikawa}
\affiliation{Technology Innovation Division, Panasonic Corporation, Osaka 571-8508, Japan}
\author{Kosei Sato}
\affiliation{Department of Engineering, Nagoya Institute of Technology, Nagoya 466-8555, Japan}
\author{Daisuke Sugita}
\affiliation{Department of Engineering, Nagoya Institute of Technology, Nagoya 466-8555, Japan}
\author{Satoshi Yotsuhashi}
\affiliation{Technology Innovation Division, Panasonic Corporation, Osaka 571-8508, Japan}
\author{Ichiro Takeuchi}
\affiliation{Department of Engineering, Nagoya Institute of Technology, Nagoya 466-8555, Japan}

\date{\today}

\begin{abstract}
Because inorganic solid electrolytes are one of the key components for application to all-solid-state batteries, high-ionic-conductivity materials must be developed. Therefore, we propose a method of efficiently evaluating the activation energy of ionic diffusion by calculating a potential-energy surface (PES), searching for the optimal diffusion path by an algorithm developed using dynamic programming (DP), and calculating the corresponding activation energy by the nudged elastic band (NEB) method.
Taking $\beta$-\ce{Li3PS4} as an example, the activation energy of Li-ion diffusion was calculated as 0.43, 0.25, and 0.40~eV in the $a$-, $b$-, and $c$-axis directions, respectively, which is in good agreement with previously reported values.
By comprehensively searching for the lowest energy path by PES-DP, the arbitrariness of the path selection can be eliminated, and the activation energy must only be calculated using the NEB method a few times, which greatly reduces the computational cost required for evaluating activation energy and enables the high-throughput screening of solid state electrolytes.
\end{abstract}

\maketitle

\section{\label{sec:intro}INTRODUCTION}
All-solid-state batteries (ASSBs), which consist of inorganic materials, are increasingly attracting attention as next-generation energy-storage devices because they are safer than organic-electrolyte-based batteries and because they potentially show higher capacities and faster charging. 
Because the solid electrolyte responsible for Li-ion conduction is one of the key materials required for fabricating ASSBs, improving the ionic conductivity of the solid electrolyte enables ASSBs to be rapidly charged and promotes their widespread application. 
Therefore, it is very important to develop high-ionic-conductivity solid-electrolyte materials.

Recently, some promising materials have been developed; for example, \ce{Li7P3S11}~\cite{seino2014_LPS}, \ce{Li10GeP2S12}~\cite{kamaya2011_LGPS}, \ce{Li7La3Zr2O12}~\cite{bernuy2014_LLZ}, \ce{Li3YBr6}~\cite{asano2018_LYB}, etc.~\cite{el2020_ternary,kim2019_borohydride,bates1993_LiPON,zhao2012_oxyhal}.
Among these materials, sulfides show very high Li ionic conductivities, exceeding those of organic-liquid electrolytes (\textit{ca}.~10~mS/cm) at room temperature.
However, sulfides show a narrow potential window~\cite{han2016_window_calc,richards2016_interface}, and their low electrochemical stabilities often degrades batteries.
However, although oxides and halides show higher electrochemical stabilities~\cite{asano2018_LYB, richards2016_interface, ohta2011_llz}, their ionic conductivities are only approximately 1~mS/cm, which is inferior to those of organic-liquid electrolytes.
Because no solid electrolyte shows both high electrochemical stability and ionic conductivity yet, the development of next-generation superionic Li conductors remains a key issue.

However, the experimental discovery of materials by trial and error alone is very time consuming.
Owing to the rapid increase in computational resources in recent years, screening using computational science has become more important~\cite{ong2013_sc,zhao2018_sc,zhu2017_sc,he2019_sc}.
The nudged elastic band (NEB) method is widely used to calculate the activation energy of Li-ion conduction, which is one of the most important properties affecting ionic conductivity.
The NEB method simulates intersite hopping of conducting atoms that are part of the diffusion path to the adjacent unit cell and obtains their activation energies.
However, the main disadvantages of the NEB method are that the beginning and end hopping points must be specified to compute the activation energy, and the method itself does not indicate the sites through which the atoms may diffuse.
Usually, considerable effort is required to find the lowest-energy diffusion path that takes the diffusing atoms to the adjacent unit cell because there are many candidates for intersite hopping routes within the crystal structure.
Although \textit{ab-initio} molecular dynamics (AIMD) calculations are sometimes used to identify the diffusion paths~\cite{mo2012_path}, they require enormous computational costs to obtain reliable results, thereby preventing high-throughput computation. 

Several methods using classical potentials instead of density functional theory (DFT)-based ones reportedly accelerate the calculation of the diffusion path and its corresponding activation energy.
Adams \textit{et al. }developed the bond valence site energy (BVSE) method~\cite{adams2000_bvse,adams2009_bvse,adams2011_bvse,chen2017_bvse,chen2019_softbv} and investigated the activation energies of various ionic conductors~\cite{zhang2020_database}.
They found that with appropriate parameter tuning, the magnitudes of the activation energies usually were consistent with the experimental values previously reported for various Li-ion conductors.
Pan \textit{et al. }combined geometrical analysis and the BVSE method to predict the diffusion path and corresponding activation energy of $\alpha$-, $\beta$-, and $\gamma$-\ce{Li3PS4}~\cite{pan2019_bvse}.
Their obtained diffusion paths agreed with those calculated based on first principles, as reported in Ref.~\cite{lepley2013_neb}.
However, calculation accuracy depends on the quality of the force fields, and it is difficult to generate a potential applicable to any element, which may limit the search for new materials in the vast elemental space.
Again, although AIMD calculations are accurate enough for most elements, the calculation cost is very high. Hence, a general robust method of quickly and accurately predicting diffusion paths for any element is required.

Therefore, we propose an efficient method of automatically predicting the diffusion paths of Li ions and corresponding activation energies by calculating the potential energy surfaces (PESs) of Li ions, searching for diffusion paths by dynamic programming (DP), and calculating the activation energy of each intersite hopping route within a diffusion path by the NEB method.
Because of the low computational cost of using our method to determine the optimal diffusion path, the activation energy can be calculated relatively quickly even with a DFT that is robust with respect to element selection. 
Our method is based on the work of Toyoura \textit{et al.}, who developed a similar technique for proton-conducting oxides~\cite{toyoura2016_dp1,kanamori2016_dp2}.
However, when their proposed method is applied to Li-ion diffusion, unlike proton diffusion, there are several problems to solve, as explained in Section~\ref{sec:PES}. We show that such problems can be solved simply by removing the Li ion nearest to the grid point.
The remainder of the paper is organized as follows. In the next section, we describe the details of our three-step method. In Section~\ref{sec:results}, we show the validity of this method and its superiority to AIMD calculations. Finally, we conclude by summarizing the main findings and their significance in Section~\ref{sec:conclusion}.

\section{\label{sec:method}METHODS}
\subsection{\label{sec:path_detection}Diffusion path detection}
\subsubsection{\label{sec:PES}PES calculation}
\begin{figure}
\centering
\includegraphics[width=\linewidth]{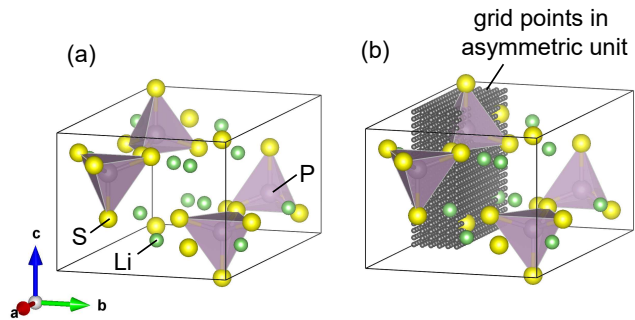}
\caption{\label{fig:asym} (a) Crystal structure of $\beta$-\ce{Li3PS4} rendered by VESTA~\cite{momma2011_vesta}. Li, P, and S atoms and \ce{PS4} polyhedra are green, purple, yellow, and purple, respectively. (b) Asymmetric unit in unit cell. Grid points in asymmetric unit are represented by black spheres.}
\end{figure}

\begin{figure}
\centering
\includegraphics[width=\linewidth]{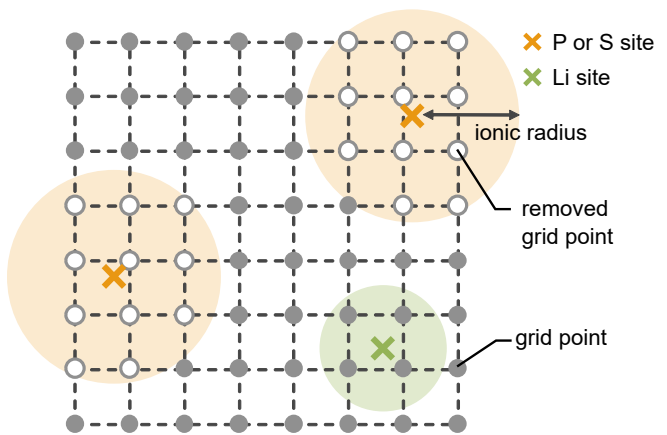}
\caption{\label{fig:within-radii} Reduction in number of grid points. Crosses represent [P or S] (orange) and Li (green) sites. Gray circles represent grid points from which energy is calculated. White circles are located within ionic radii of P or S sites and are removed from calculation target.}
\end{figure}

Specifically, $\beta$-\ce{Li3PS4} (ID: mp-1097036, space group: $Pnma$), as obtained from the materials project database~\cite{jain2013_mp}, was used as a model structure.
TABLE~\ref{tab:lattice} shows the lattice parameters of $\beta$-\ce{Li3PS4}.
\begin{table}[b]
\caption{\label{tab:lattice} Lattice parameters of $\beta$-\ce{Li3PS4}.}
\begin{ruledtabular}
\begin{tabular}{cccccc}
 & & \AA & & degrees\\
\colrule
Lattice & $a$ & 13.42 & $\alpha$ & 90.00\\
parameters & $b$ & 7.924 & $\beta$ & 90.00\\
 & $c$ & 6.221 & $\gamma$ & 90.00\\
\colrule
\textrm{Site}&
\textrm{Wyckoff}&
\textrm{$x$}&
\textrm{$y$}&
\textrm{$z$}&
\textrm{Occupancy} \rule[0mm]{0mm}{5mm}\\
\colrule
Li1 & 8d & 0.1594 & 0.0004580 & 0.7989 & 1\\
Li2 & 4c & 0.06986 & 0.2500 & 0.2116 & 1\\
P1 & 4c & 0.09252 & 0.7500 & 0.3042 & 1\\
S1 & 8d & 0.1644 & 0.5364 & 0.1974 & 1\\
S2 & 4c & 0.05266 & 0.2500 & 0.7999 & 1\\
S3 & 4c & 0.09559 & 0.7500 & 0.6371 & 1\\
\end{tabular}
\end{ruledtabular}
\end{table}
To obtain PES, we introduced fine ($40 \times 40 \times 20 = 32,000$) grid points at an interval of approximately 0.3~\AA~in the unit cell and calculated the potential energy when an additional Li ion was placed on a grid point.
To reduce the calculation cost, only the grid points within the asymmetric unit were evaluated.
In $\beta$-\ce{Li3PS4}, the asymmetric unit satisfying [$0 \leq x \leq 0.5,  0 \leq y \leq 0.25,  0 \leq z \leq 1.0$] contains 2,520 grid points (Fig.~\ref{fig:asym}), where $x$, $y$, and $z$ denote the three-dimensional fractional coordinates of the structure.
In addition, the grid points that overlap the \ce{P^5+} and \ce{S^2-} ions (that is, where the minimum distance to the center of these ions is shorter than each ionic radius: 0.17~\AA~for \ce{P^5+} and 1.84~\AA~for \ce{S^2-}) were omitted from the calculation (Fig.~\ref{fig:within-radii}) because the potential energy of these grid points is expected to be exceedingly high for Li diffusion, and the diffusion path will not include such grid points.
Finally, the number of grid points from which energy is calculated was reduced to 1,009.
The validity of the ionic-radius-based reduction in the number of grid points is confirmed in Section~\ref{sec:calc_num}.

\begin{figure}
\centering
\includegraphics[width=\linewidth]{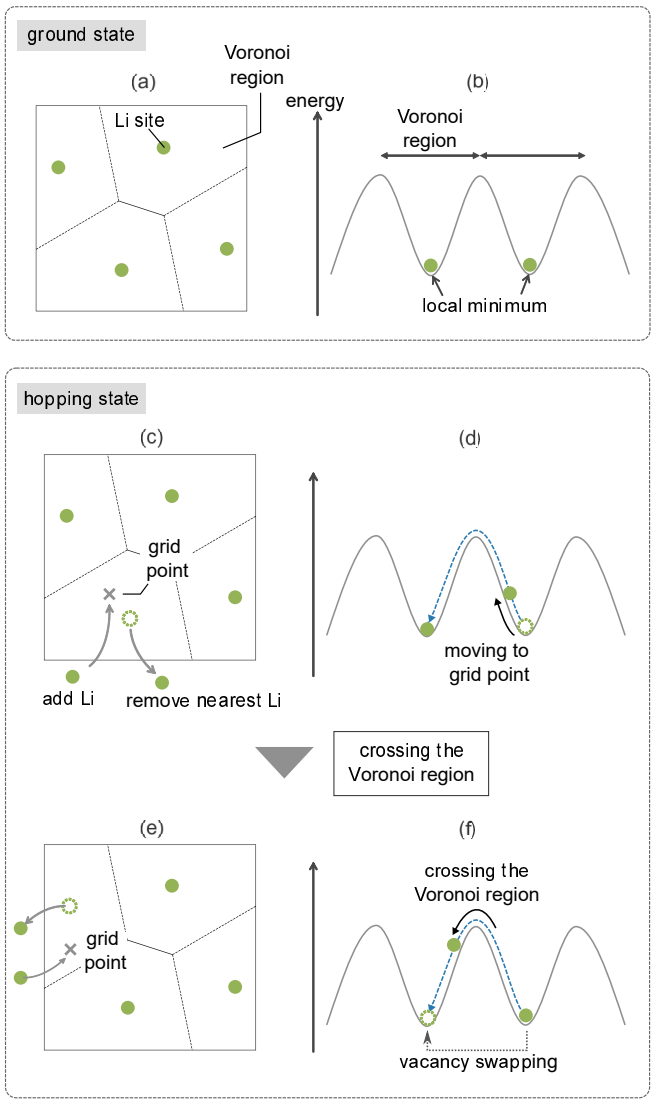}
\caption{\label{fig:voronoi} Voronoi tessellation of Li sublattice and PES in Voronoi region. Li sites at Voronoi center correspond to local minimum of PES\@. When additional Li ion is introduced at grid point in Voronoi region, one Li ion at center of Voronoi region is removed. (a, b) Ground state, (c)--(f) hopping state. After crossing Voronoi region, vacancy moves to adjacent Li site because Li site nearest to grid point changes.}
\end{figure}

We introduced an additional Li ion at each grid point in the unit cell and obtained a list of potential energies by computing the energy according to single-point DFT calculations (\textit{i.e}.~without structural optimization).
The introduction of an additional Li ion causes extra Li–Li interactions and lattice expansion leading to the inaccurate evaluation of PES. 
To minimize this effect, we removed from the unit cell the Li ion nearest to the added one, which shows the strongest Li–Li interaction, keeping the charge-neutrality of the unit cell (Fig.~\ref{fig:voronoi}).
This procedure is suitable for reproducing the hopping conduction mechanism.
To explain why, it is convenient to consider the Voronoi tessellation of the Li sublattice before an additional Li ion is introduced.
Then, Voronoi centers correspond to Li sites in the unit cell, meaning that Voronoi centers correspond to the local minimum of the PES; that is, the endpoints of the Li hopping conduction.
When climbing the PES, as shown in Figs.~\ref{fig:voronoi}(c) and (d), the starting point of the Li hopping should become a vacancy, which can be expressed by Li ions moving from a Li site to a grid point.
After the Li ion crosses the Voronoi boundary, as shown in Figs.~\ref{fig:voronoi}(e) and (f), the nearest Li site changes to the neighbor Voronoi center, and the hopping endpoint becomes a vacancy.
Such vacancy swapping meets the hopping conduction requirement that the endpoints must be vacancies.

Incidentally, it is unnecessary to carry out the procedure for proton-conducting oxides, for which it has been previously proposed that proton diffusion proceeds by the Grotthuss mechanism~\cite{agmon1995_grotthuss,nowick1995_oxy_gro}.
Protons do not hop between specific lattice sites, rather they move between adjacent oxygen atoms by hydrogen bonding.
Conducting protons are taken up into the structure by \ce{H2O} molecules trapped in oxygen vacancies.
Because protons do not occupy any sites in the unit-cell crystal but are externally doped in low concentrations, it is unnecessary to consider proton interactions. Moreover because the proton ionic radius is negligible compared to the lattice size, lattice-expansion-induced energy change is unlikely to occur.

Energies were calculated using DFT according to the projector-augmented wave method~\cite{blochl1994_paw} implemented in the Vienna \textit{ab-initio} simulation package (VASP)~\cite{kresse1996_vasp}.
The Perdew–Burke–Ernzerhof functional was used as the exchange-correlation functional under generalized gradient approximation ~\cite{perdew1996_pbe}.
The plane-wave cutoff energy was set to 520~eV, and 2×4×5 $k$-point sampling was used.
We computed the potential energy by a single point-calculation without structural optimization, which would have caused the Li ion at the grid point to be relaxed to the nearest Li site defined as a vacancy site in our model.

\subsubsection{\label{sec:DP}Optimal path search by DP}
DP is a framework that divides a target problem into multiple subproblems and solves them while effectively using their solutions properly stored in memory~\cite{bellman1952theory, denardo2012dynamic}.
The optimal path search is one of the problems that can be solved efficiently using DP\@. 
We obtained the global minimum, bottleneck point, and diffusion path from the PES by the previously developed DP-based optimal path-search algorithm~\cite{toyoura2016_dp1,kanamori2016_dp2}, which finds the path showing the lowest diffusion cost starting from the global minimum in a unit cell to the corresponding point in the adjacent one. 
By finding three linearly independent paths along the crystal axis, any three-dimensional path can be represented.
Because the activation energy differs depending on the diffusion direction in an anisotropic crystal, we computed the path showing the lowest activation energy for each direction.

\subsection{\label{sec:NEB}Activation energy calculated by DFT-based NEB method}
After the activation energy of the diffusion path was calculated by the method described in Section~\ref{sec:DP}, the NEB method was used to evaluate the energy more accurately. The activation energies of the intersite hopping routes (part of the diffusion path) connecting the two Li sites in the unit cell were calculated assuming a single vacancy diffusion mechanism. The force convergence conditions were set to 0.02~eV/{\AA} to optimize the endpoint structures and 0.05~eV/{\AA} for the NEB calculation. The plane-wave cutoff energy was set to 520~eV, and 1×3×4 $k$-point sampling was used. 
Using \texttt{pymatgen}~\cite{ong2013_pymatgen}, the reaction coordinates were normalized from 0 to 1, and the energy curve was obtained by cubic spline interpolation.

\section{\label{sec:results}RESULTS AND DISCUSSION}
\subsection{\label{sec:lps_results}Diffusion path and activation energy}
\begin{figure*}
\centering
\includegraphics[width=\linewidth]{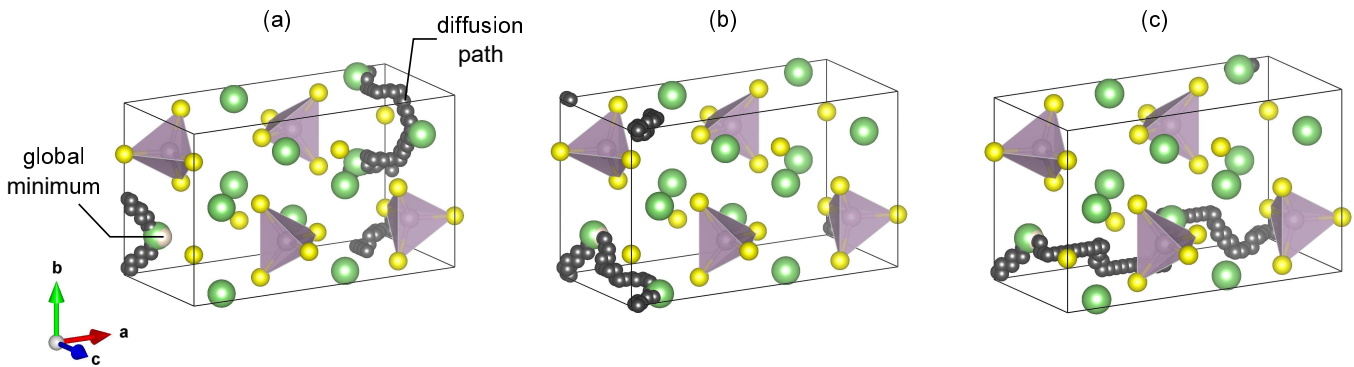}
\caption{\label{fig:DP-path} Lowest-activation-energy diffusion paths calculated by DP\@ in each crystal-axis direction. Diffusion paths to (a) $b$, (b) $c$, and (c) $a$ axes are shown by black spheres, and white spheres represent PES\@ global minimum.}
\end{figure*}

The PES of the Li-ion diffusion was evaluated by calculating the energy of the $\beta$-\ce{Li3PS4} structure, as explained in detail in Section \ref{sec:PES}. Figures~\ref{fig:DP-path}(a)–\ref{fig:DP-path}(c) show the three paths obtained along the crystal axes by applying DP to the PES. 
\begin{figure}
\centering
\includegraphics[width=\linewidth]{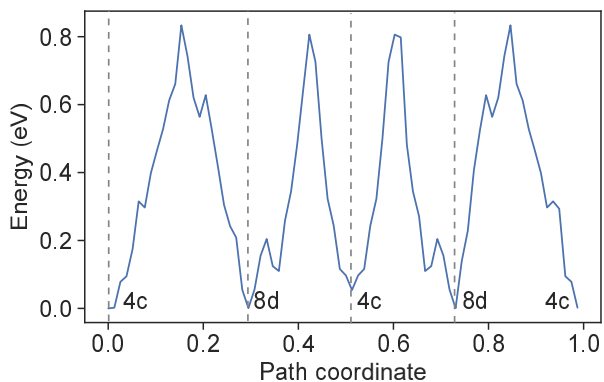}
\caption{\label{fig:LPS_PES} PES along $b$-axis diffusion path. Dotted line corresponds to grid point nearest to Li site.}
\end{figure}
The path showing the lowest activation energy was in the $b$-axis direction followed by the $c$- and $a$-axis directions, which is consistent with the results reported by Pan \textit{et al. }\cite{pan2019_bvse}.
Along all of these paths, the diffusing Li ions pass through the original unit-cell Li sites while avoiding the \ce{PS4} polyhedra.
To investigate the cell size effect on the diffusion path, we have repeated our path search using 2 × 2 × 2 supercell, obtaining almost identical paths to the unit-cell case. Therefore, we will limit our discussion using unit cell hereafter.
Figure~\ref{fig:LPS_PES} shows the PES of the path along $b$-axis direction. 
While the Li ions were diffusing to the adjacent unit cell, four local minima were observed at path coordinates 0.00, 0.29, 0.51, and 0.73, corresponding to the original Li sites and indicating that hopping conduction between Li sites could be reproduced by the calculated PES and DP analysis.
Along this diffusion path, the activation energy of the Li-ion diffusion was calculated as 0.84~eV, which is considerably higher than previously reported activation energies (0.16~eV~\cite{homma2011_lps} and 0.24~eV~\cite{stoffler2018_lps} in experiments and 0.4~eV~\cite{de2018_lps} in AIMD simulations) because the unit cell was not optimized for calculating the PES\@.
In each structure used to calculate the PES, atomic positions were fixed at their original positions except for the added and removed Li ions.
Because the most stable positions of the atoms change owing to the introduction of an additional Li ion to a grid point, the static calculation overestimates the potential energy.
However, removing the Li ion nearest to the grid point minimizes the nearest Li-Li repulsion force and the change in cell size.
Therefore, the relative activation-energy trend was not significantly affected, and a reasonable diffusion path, as shown in Fig.~\ref{fig:DP-path} was obtained.

\begin{figure}
\centering
\includegraphics[width=\linewidth]{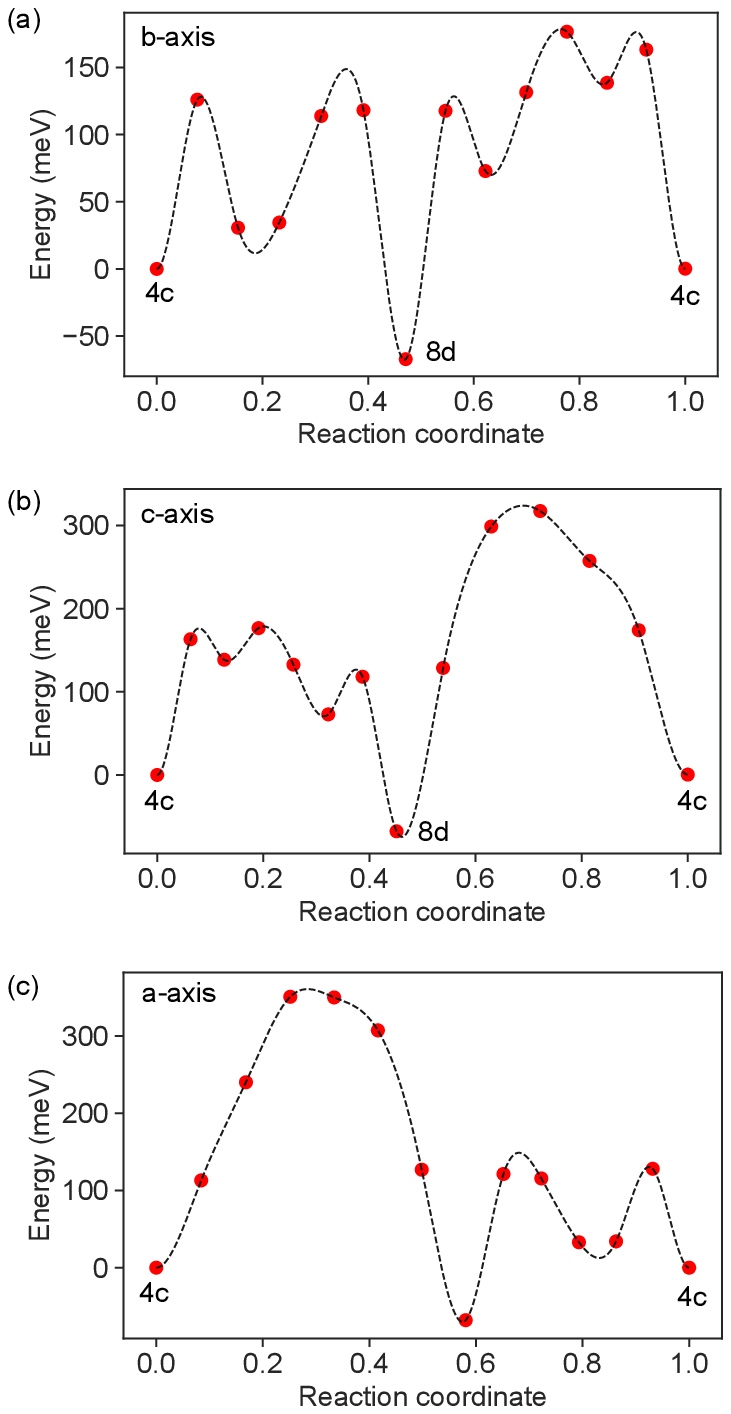}
\caption{\label{fig:LPS_NEB} Energy profiles of diffusion path along each crystal axis, as calculated by NEB method. Reaction coordinates are normalized from 0 to 1, and red circles correspond to each NEB image. Dashed line indicates energy curve obtained by cubic spline interpolation.}
\end{figure}

\begin{figure}
\centering
\includegraphics{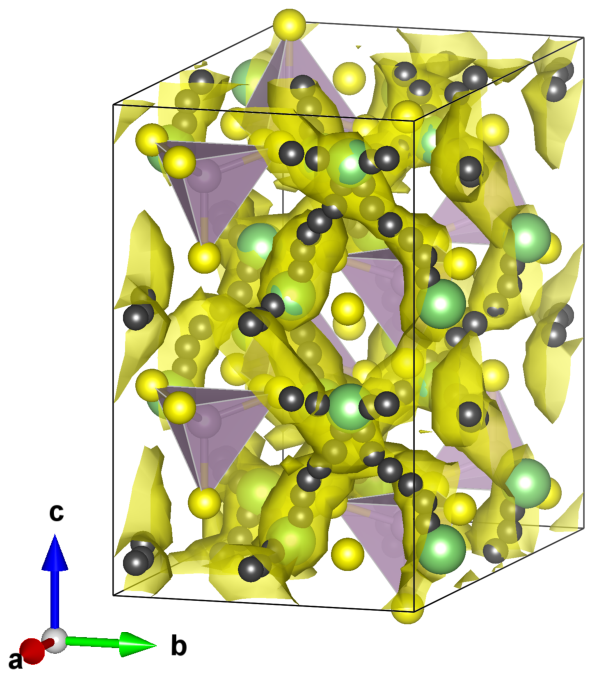}
\caption{\label{fig:MD_vs_NEB} Diffusion paths calculated by AIMD and NEB methods. Yellow surface indicates region showing high probability density of Li ions, as calculated by AIMD, and black spheres show NEB-optimized diffusion paths.}
\end{figure}

To improve the accuracy of the activation-energy evaluation, the activation energies of the intersite hopping routes, which are part of the diffusion path, were calculated by the NEB method (Fig.~\ref{fig:LPS_NEB}).
For the paths along the $b$- and $a$-axis directions, only half of the entire path must be calculated because of the symmetry of the crystal structure. That is, the reaction coordinates from 0 to 1, as shown in Figs.~\ref{fig:LPS_NEB}(a) and \ref{fig:LPS_NEB}(c), correspond to the path coordinates from 0 to 0.5, as shown in Fig.~\ref{fig:LPS_PES}.
Table~\ref{tab:Ea} shows the activation energies obtained by the PES-DP and NEB methods for each path along the crystal axes.
\begin{table}[b]
\caption{\label{tab:Ea} Activation energies obtained by PES-DP and NEB methods for paths along crystal axes.}
\begin{ruledtabular}
\begin{tabular}{cccc}
  & \multicolumn{3}{c}{Activation energy (eV)} \\
\cline{2-4}
 Method & $a$-axis & $b$-axis & $c$-axis \\
\colrule
 PES + DP & 1.09 & 0.84 & 1.03 \\
 NEB & 0.43 & 0.25 & 0.40 \\
\end{tabular}
\end{ruledtabular}
\end{table}
The activation energies calculated from the NEB method were 0.25, 0.40, and 0.43~eV for the $b$-, $c$-, and $a$-axis directions, respectively, which were on the same sequence as the activation energies calculated from the PES-DP method.
The activation energy in the $b$-axis direction was the lowest and was close to previously reported experimental~\cite{homma2011_lps, stoffler2018_lps} and AIMD-calculated activation energies~\cite{de2018_lps}.
The NEB calculation converged after relatively few (10--100) self-consistent field (SCF) calculations, probably because of the reasonable initial path.
We then performed the gamma-point-only AIMD calculation to determine whether the NEB-obtained diffusion path was accurate.
Figure~\ref{fig:MD_vs_NEB} shows the result obtained by comparing the AIMD-calculated and NEB-optimized paths.
The path obtained by the NEB method was extended to the entire crystal structure by symmetry operations.
The yellow surface within which the Li ion shows a high AIMD-calculated probability density is consistent with the NEB-optimized path shown by the black spheres.
Therefore, the PES-DP-calculated path can be a good initial estimate for the actual Li-ion diffusion path.
However, the energy profiles between the hopping sites are slightly different.
In the calculated PES (Fig.~\ref{fig:LPS_PES}), only one bottleneck (0.15) was observed between the 4$c$-8$d$ sites (in the path-coordinate range 0.00--0.29), whereas in the NEB results [Fig.~\ref{fig:LPS_NEB}(a)], one local minimum (0.20) and two bottlenecks (0.10 and 0.37) were observed in the corresponding region (in the reaction-coordinate range 0.00--0.47).
Although there are subtle differences in energy transitions, the intersite hopping route showing the lowest activation energy was correctly identified.

\begin{figure}
\centering
\includegraphics[width=\linewidth]{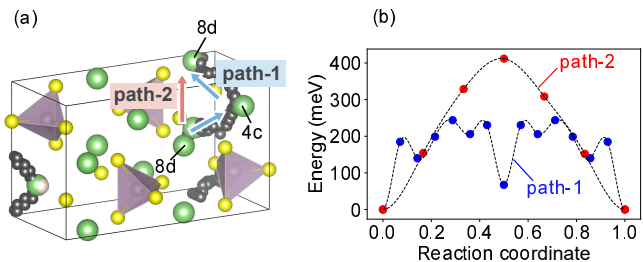}
\caption{\label{fig:8d_site_path} (a) Two diffusion paths between 8$d$ Li sites. Path obtained by DP is 8$d$-4$c$-8$d$ (path 1), and path connecting shortest distance is 8$d$-8$d$ (path 2). (b) Energy profiles for Li-ion diffusions of paths 1 and 2.}
\end{figure}

Our PES-DP method eliminates the arbitrariness of selecting the intersite hopping route for calculating activation energies and can automatically find the diffusion path showing the lowest activation energy.
Interestingly, our PES-DP method found a nontrivial path rather than a path connecting the shortest distances.
As shown in Fig.~\ref{fig:8d_site_path}(a), there are two possible intersite hopping routes between 8$d$ Li sites along the $b$-axis (i.e., the 8$d$-4$c$-8$d$ and 8$d$-8$d$ paths).
Figure~\ref{fig:8d_site_path}(b) shows the activation energy calculated by the NEB method for these two paths.
Our method predicted the 8$d$-4$c$-8$d$ path (path 1) showing an activation energy of 0.25~eV [Fig.~\ref{fig:LPS_NEB}(a)], while the 8$d$-8$d$ path (path 2) showed a higher activation energy of 0.41~eV.
Therefore, our method can correctly detect the entire diffusion path including the sites through which the Li ions pass.\footnote{Experimentally, the $\beta$-\ce{Li3PS4} is identified to occupy both 4$b$ and 4$c$ sites with fractional occupancies~\cite{homma2011_lps}. Because we have used the structure from the Materials Project database, in which Li atoms are idealized not to occupy 4$b$ sites, the diffusion paths do not pass 4$b$ sites. In fact, when we performed the PES calculation using the idealized structure with the occupancy of the Li 4$b$ sites unity and that of the Li 4$c$ sites zero~\cite{lepley2013_neb}, the path with the lowest activation energy was found to contain the 4$b$ sites.}
When calculating the activation energy of the intersite hopping by the NEB method, the diffusion path is assumed to be the shortest path between the sites. However, our results indicate that the detoured path (8$d$-4$c$-8$d$), rather than the shortest one, shows a lower activation energy. 
Therefore, if only the shortest path is calculated according to the general method, an inaccurate activation energy may be obtained.
Our PES-DP method automatically and efficiently finds diffusion-path candidates and promotes high-throughput activation-energy calculations, which are essential for discovering high-ionic-conductivity materials.

\subsection{\label{sec:calc_num}Fewer grid points required for PES calculations}
\begin{figure}
\centering
\includegraphics[width=\linewidth]{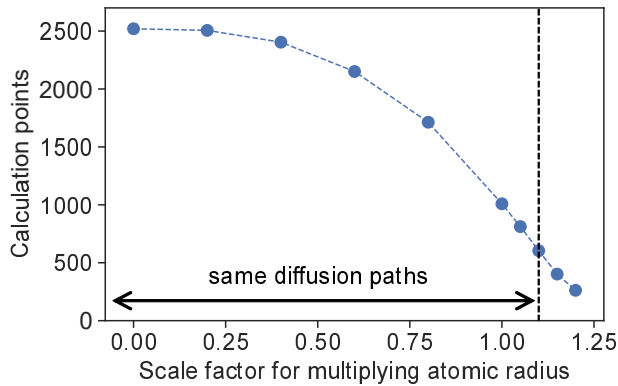}
\caption{\label{fig:calc_num} Relationship between number of calculation points required and ionic-radius scale factor. Grid points within scaled-ionic-radius (\textit{i.e}., ionic radius multiplied by abscissa) distance from atomic coordinates of P or S ions are omitted from calculations.}
\end{figure}

As described in Sec.~\ref{sec:PES}, we did not use the grid points that overlapped the \ce{P^5+} and \ce{S^2-} ions (i.e., within the ionic radii of \ce{P^5+} and \ce{S^2-}) to evaluate the PES.
To justify following such a procedure, we compared the diffusion path obtained in Sec.~\ref{sec:lps_results} with that obtained when calculating all the grid points in the asymmetric unit.
We also investigated the number of calculations required owing to ionic-radius scaling.
When a Li ion is at a grid point within the ionic radius of each atom, the potential energy will be extraordinarily high, and its grid point cannot be a diffusion path.
Therefore, we set the energy of such grid points to a positive value (\textit{e.g}., 1.0~eV), which corresponds to assigning unusually large energy barriers and diffusion costs, to find the diffusion path.
Figure~\ref{fig:calc_num} shows the change in the number of calculations (\textit{i.e}., number of grid points) required as the ionic radius is scaled and grid points within the scaled ionic radius are removed.
The abscissa and ordinate denote the ionic-radius scale factor and the change in the number of grid points required as the ionic radius changes, respectively.
The number of grid points required decreased dramatically with increasing scale factor.
Up to a scale factor of 1.1, the diffusion path was exactly identical as that obtained by calculating all the grid points. 
As explained in Sec.~\ref{sec:PES}, we used a scale factor of 1.0, and the number of calculations required decreased from 2,520 to 1,009.
Although such methods should be adopted to reduce computational costs, the scale factor should be carefully chosen depending on the material system.
Because sulfides show high polarizability, the ionic radius (\textit{i.e}., spread of the S-ion electron cloud) easily changes during Li-ion diffusion, meaning that with decreasing S-ion radius, the scale factor should be decreased to avoid underestimating the number of grid points required for PES calculations.
We expect that a scale factor of 1.0 can be used when calculating the PES of materials that are more ionic than $\beta$-\ce{Li3PS4}, such as oxides and halides.

Because few grid points are required using our method of finding the diffusion path, calculation time can be considerably reduced.
AIMD calculations perform $O(10^4-10^5)$ SCF calculations to obtain the diffusion path, because Li hopping rarely occurs at moderate temperatures~\cite{he2018_statistical}.
Therefore, AIMD calculations are very inefficient for detecting the diffusion path.
Our method, on the other hand, requires only $O(10^2-10^3)$ SCF calculations to detect the path and $O(10^2)$ to compute the activation energy by the NEB method.
Moreover, because our method is based on first-principles calculations, it is very robust compared to classical force-field-based methods such as the BVSE method mentioned in Sec.~\ref{sec:intro}. Therefore, our method is sufficiently accurate that it can be applied to various materials.

\begin{figure}
\centering
\includegraphics{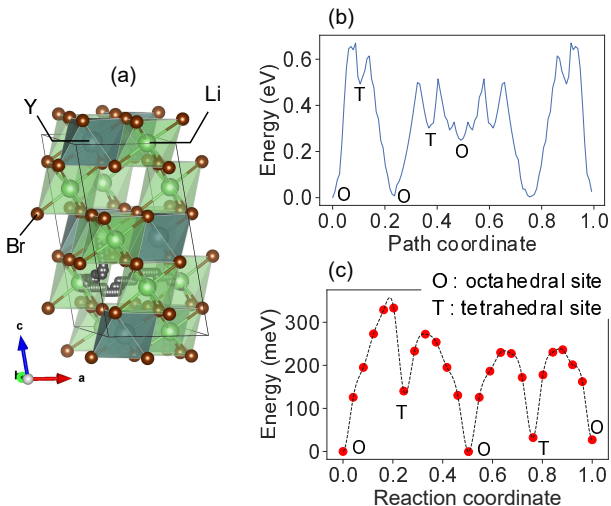}
\caption{\label{fig:LYB} (a) Diffusion path of \ce{Li3YBr6} showing lowest activation energy and (b) corresponding PES. (c) Energy profile of diffusion path calculated using NEB method. O and T indicate octahedral and tetrahedral sites, respectively. Reaction coordinates from 0 to 1 in panel (c) correspond to path coordinates from 0 to 0.5 in panel (b).}
\end{figure}

\subsection{\label{sec:lyb}Application to \ce{Li3YBr6}}
To support the generality of our method, we calculated the Li-ion diffusion path of a halide material (\ce{Li3YBr6}), which has recently attracted considerable attention.
Halides are suitable for verifying the generality of our method because they show more ionic bonds than sulfides, which show more covalent ones.
Previously, it has been reported that \ce{Li3YBr6} shows a fcc anionic sublattice wherein Li cations partially occupy octahedral sites~\cite{asano2018_LYB}.
Because the structure showed a $C2/m$ space group, grid points indicated by [$0 \leq x \leq 0.25,  0 \leq y \leq 0.25,  0 \leq z \leq 0.25$] were used in the asymmetric unit.
Figures~\ref{fig:LYB}(a) and \ref{fig:LYB}(b) display the path showing the lowest activation energy, as found by DP and PES.
Similar to the $\beta$-\ce{Li3PS4} results, although the activation energy (0.67~eV) was considerably higher than the experimental (0.37~eV~\cite{asano2018_LYB}) and AIMD-calculated ($0.28 \pm{0.02}$~eV~\cite{wang2019_lyb}) values, the NEB method provided more accurate activation energies.
The NEB result was 0.36~eV, as shown in Fig.~\ref{fig:LYB}(c), which is in agreement with the previously reported values.
In the obtained diffusion path, the Li ions hop between the two octahedral Li sites through the tetrahedral vacancy sites, which is consistent with previously reported AIMD calculation results~\cite{wang2019_lyb}. 

Here, we comment on the ``pinball'' model recently proposed by Kahle \textit{et al. }to calculate efficiently the PES and dynamics of Li ions in a solid-state ionic conductor~\cite{kahle2018modeling}. This model assumes that the Li atoms to be fully ionized, and that the positions and valence charge density of host lattice atoms be frozen. Because the self-consistent calculation is required only for the initial step and the dynamics is computed for the Li ``pinballs'' alone, the diffusion properties of Li ions can be computed very efficiently. The model has reproduced the results of AIMD calculations for oxides, but not as well for less rigid materials like sulfides, because it does not include lattice vibration at present. By contrast, our method accounts for the lattice movement by the NEB calculation after the path search, and in fact, has succeeded in computing the activation energies for sulfides and chlorides. 

\section{\label{sec:conclusion}CONCLUSION}
We showed that Li-ion diffusion paths could be efficiently determined by calculating the PES and applying a DP-based path-search algorithm. 
We tried to express the hopping conduction of Li intersites by introducing fine grid points and removing the Li ion nearest to the additional Li ion inserted at the grid point to calculate PES\@.
The obtained optimal diffusion path of $\beta$-\ce{Li3PS4} was along the $b$-axis direction, and the corresponding activation energy calculated by the NEB method was in good agreement with the experimental value.
Because our method is based on DFT calculations, it is sufficiently accurate and more robust than classical force-field-based methods.
Although our method requires a relatively large number of calculations for crystal structures with low symmetry, it can be expected that the computational cost can be reduced by PES evaluation using the Gaussian process and Bayesian optimization~\cite{toyoura2016_dp1,kanamori2016_dp2}.
In this paper, we assume a single-atom diffusion model, but in some of the superionic conductors, the concerted conduction, in which multiple atoms move simultaneously, can be important~\cite{he2017origin}. We plan to work on such a mechanism by extending our path search method in the future.
Our automated and non arbitrary path detection method greatly reduces the computational cost for evaluating activation energies of a diffusing ion and enables the high-throughput screening of solid state electrolytes.

\section{\label{sec:acknowledgments}ACKNOWLEDGMENTS}
The authors sincerely appreciate S. Hasegawa, A. Sakai, T. Asano and T.Yokoyama of the Panasonic Corporation for fruitful discussions.
We would like to thank Editage (www.editage.com) for English language editing.

\end{document}